\documentclass[lettersize,journal]{IEEEtran}
\usepackage{amsmath,amsfonts}
\usepackage{algorithmic}
\usepackage{algorithm}
\usepackage{array}
\usepackage{textcomp}
\usepackage{stfloats}
\usepackage{url}
\usepackage{verbatim}
\usepackage{graphicx}
\usepackage{cite}
\usepackage{xcolor}
\usepackage{booktabs, multirow} 
\usepackage{soul}
\usepackage[table]{xcolor} 
\usepackage{changepage,threeparttable} 
\usepackage{overpic}
\usepackage{caption}
\usepackage{subcaption}
\usepackage{booktabs, multirow} 
\usepackage[super]{nth}

\newcommand{\review}[1]{\textcolor{black}{#1}}

\begin{document}

\title{Complete Integration of Team Project-based Learning into a Database Syllabus}

\author{
Sergio Iserte, 
Vicente R. Tomás,
Miguel Pérez,
Maribel Castillo,
Pablo Boronat,
Luis A. García
\\
\vspace{0.3cm}
\textbf{Universitat Jaume I, Spain}
\thanks{Manuscript received April 19, 2021; revised August 16, 2021.}}

\markboth{Journal of \LaTeX\ Class Files,~Vol.~14, No.~8, August~2021}%
{Shell \MakeLowercase{\textit{et al.}}: A Sample Article Using IEEEtran.cls for IEEE Journals}

\IEEEpubid{0000--0000/00\$00.00~\copyright~2021 IEEE}

\maketitle

\begin{abstract}
Team project-based learning (TPBL) combines two learning techniques: project-based learning (PBL) and teamwork. This combination leverages the learning outcomes of both methods and places students in a real work situation where they must develop and solve a real project while working as a team. TPBL has been used in two advanced database subjects in Jaume I University (UJI)'s Computer Science degree program. This learning method was used for four years (academic years from 2018/19 to 2021/2022) with positive outcomes. This study presents the project development, which includes teamwork formation, activities, timetable, and exercised learning competencies (both soft and specific). Further, the project's results were evaluated from three different perspectives: a) teamwork evaluation by teammates, b) Students’ opinions on the subject and project, and c) subject final grades.    
\end{abstract}


\section{Introduction}

\IEEEPARstart{T}{eam} project-based learning (TPBL) combines teamwork and project-based Learning (PBL) methods. PBL is based on a project as the main axis of the learning process~\cite{Pearlman2000}, 
which guides students through the development of a subject by posing a challenge that cannot be solved solely by applying superficial knowledge. 
In addition, PBL encourages an active and motivated student attitude since they must take decisions when proposing and defending solutions.
The basic idea of PBL is to place students in real situations that require them to analyze, design, implement and evaluate projects that have real applications, beyond the classroom examples. In this way, students will be able to apply what they have learned to solve problems in their future professional activities or propose improvements in the communities where they work.

PBL has been applied as a teaching technique in several disciplines and learning levels. It has been widely used in higher education levels~\cite{GUO2020101586,10481/53934,KUO20191}, including Computer Science degree programs~\cite{connolly2006constructivist,Nattassha2015DatabaseAA,Garcia-Robles,Caceffo}.

Teamwork is extremely effective and appropriate in the classroom, allowing students to not only expand their knowledge and retain  concepts  and information over time but also develop communication skills~\cite{Oakley04turningstudent, Sulaiman2017, Pitsoe2014}.
Particularly, teamwork in a project is considered an essential skill to be developed during the education of new professionals~\cite{Berry2001,De_Prada}.
The combination of PBL and teamwork allows the active development of specific and basic skills such as the collection and interpretation of information, the issuance expression of opinions and judgments, and the presentation of arguments, thus promoting the transfer of ideas and consensus.

This study not only describes the TPBL method but also investigates how it was implemented in two subjects on advanced databases at Jaume~I University (UJI)'s Computer Engineering degree. 
In these subjects, the design of information systems is studied in-depth by applying relational databases.
The specific competencies of the subjects focus on the definition, design, implementation and maintenance of information systems, and providing solutions to integrate problems based on Information Technology (IT). 

The aforementioned specific competencies are directly related to the knowledge of real-world information systems. However, combining PBL with teamwork facilitates the achievement of learning results and encourages the acquisition of basic skills such as problem-solving, leadership, and critical reasoning.
In summary, TPBL is a technique that is well suited to competencies addressed in these subjects.

This study presents the experience and results of the gradual implementation of the PBL combined with teamwork during the last four academic years (from 2018/19 to 2021/22), culminating in the full adoption of TPBL in the final year.
Results from three different perspectives such as a) students' peer evaluation,  b) students’ opinions on the project's impact, and c) subject final results, were analyzed.

Section~\ref{sec:background} presents the context of the project implementation.
Subsequently, Section~\ref{sec:activities} describes the project, activities, competencies, and evaluation method in detail.
Section~\ref{sec:results} analyzes and presents the results.
\review{Finally, Section~\ref{sec:conclusions} presents the conclusions of our TPBL learning experience and the issues that will be considered to continue with the presented learning method within the Educational Innovation Group, of which the signatories of this study are members.}

\section{Background}\label{sec:background}
\IEEEpubidadjcol

The TPBL methodology presented in this study is developed within the framework of the subjects \textit{Design of Database Systems} and \textit{Design and Implementation of Databases}. 
These subjects, are respectively, part of the \textit{Information Systems (IS)} and \textit{Software Engineering (SE)} specializations offered in the syllabus of the \textit{Computer Science} bachelor's degree program at UJI.
Whereas IS specialization deals with information processing for organizational procedures and systems, SE focuses on developing software solutions using advanced algorithms and data structures~\cite{CS2020}. 
Thus, databases play a crucial role in both specializations.

Both subjects are taught together because they share the same characteristics in their respective specializations, such as the same extension, academic year, and competencies. Students share classrooms, contents, lecturers, and evaluations.

These subjects are taught in three parts: theory (12 sessions of 1.5 h), problems (13 sessions of 1.5 h), and laboratory practices (9 sessions of 2 h). 
Laboratory sessions start in the \nth{6} week, once the needed theoretical contents have been introduced. 

The evaluation of the subject is divided into the following two parts:
\begin{itemize}
    \item The continuous assessment (60\%) consists of the following:
    \begin{itemize}
        \item The delivery of activities in the theory/problems sessions (20\%),
        \item The delivery of the practice bulletins at laboratory sessions (20\%), and 
        \item The development of a final team project (20\%).
    \end{itemize}
    \item A final exam (40\%).
\end{itemize}
Students must obtain at least 50\% of the grade in each of the two parts to pass the course.


During the last four years, the subject's teaching staff has worked on incorporating TPBL into the teaching methodology. 
To achieve this, the subject's teamwork was modified, and it became a larger project, in which teams had to develop different activities throughout the three parts of the classes (theory, problems and laboratory practices).
During the project implementation, the following generic and transversal competencies, which are essential to obtaining a graduate degree in Computer Engineering at UJI, were worked on:
\begin{itemize}
    \item Capacity for analysis and synthesis.
    \item Organizational and planning skills.
    \item Oral and written communication in mother-tongue language.
    \item Ability to manage information.
    \item Problem resolution.
    \item Teamwork promoting respect for diversity, equity, and gender equality.
    \item Autonomous learning. 
\end{itemize}

\section{Activities}\label{sec:activities}

\subsection{Description}

The set of activities designed to implement this potential learning method is divided into two groups. 
The first group includes simple tasks used to introduce basic database concepts. The other group involves complex activities, identified as \emph{core activities}. These activities are studied, discussed, and performed by working teams, with the proposed project being the main axis. 
Core activities refer to database design covering all levels: conceptual, logical, and physical, as well as database implementation in a specific DataBase Management System (DBMS), the creation of different business rules, and a final report which includes a backup of the database. 
The eight core activities proposed to the students are as follows:
\begin{enumerate}
    \item Description of requirements. In the first teamwork session, the teacher assumes the role of a client and describes the company to be modeled. From this point, the team analyzes the information system requirements.
    \item Conceptual design of the data model. In this part, the semantics of the company's data are defined considering the data and the different user perspectives. The physical representation is excluded at this stage. The result is a conceptual schema.
    \item Logical design of the data model. In this activity the conceptual schema is transformed into a logical one in which the data structures of the database model are identified.
    \item Physical design: designing tables. This involves implementing the database from the logical schema defined in the previous step. This new task involves defining a set of tables and their constraints and creating them in a specific DBMS.
    \item Physical design: the business rules. These rules model specific constraints on the information system for a particular company.  
    This activity involves proposing the business rules according to the project. 
    Each team member's activity is determined by this proposal.  Then, they are implemented in the database and tested together. 
    \item  Physical design: roles and views. This is the third activity related to physical design. The data model is analyzed from a security and access perspective.
    Teams must define the roles of the different users and their views to access the data. 
    \item Physical design: business rules on views. This is the last task in the design stage. Here, the business rules are implemented based on the  previously defined  views.
    \item Final delivery. This is the final task and it matches the end of the course. At that time, the teams must deliver the project in a predetermined format, including a copy of the database implementation. A template that specifies the sections and format of the report is provided.
\end{enumerate}

\subsection{Development} 
The project is divided into theory, problem, and laboratory sessions throughout the academic semester. The first three core activities (description of requirements, conceptual design, and logical design) are developed in two sessions, as follows:
\begin{enumerate}
\item The teamwork begins in a problem session. In the classroom, each team undertakes the assigned activity by resolving its doubts with the teacher, who acts as the client. This activity can be completed outside the classroom and it must be delivered two days before the next problem session. 
\item Then, in the first part of the following problem session, each team must perform a peer review of the solutions provided by \review{two other teams}. With this knowledge of alternative solutions to the task, students can incorporate some of the features of these alternative solutions into their proposals. The teachers, who have previously reviewed the delivered activities, discuss each team's proposed solution and offer modifications or alternatives.
\end{enumerate}

Once the logical design is complete, each team must execute the next activity outside the classroom. This activity includes the physical design and the tables. The resulting database must be deployed on the database server. This activity must be completed before the next problem session. 

The remaining physical design activities are distributed as follows: 
\begin{itemize}
\item Teams make a first theoretical proposal for an activity in the first related theory session. 
 \item Then, a subtask of the theoretical proposal is assigned to each team member. 
 \item Each team member implements his/her assigned subtask in the corresponding laboratory sessions. 
 \item Once the laboratory practice is completed, teams integrate all subtasks of the core activity and implement them into their database.
\end{itemize}

The final project delivery, which is the last core activity, is performed outside the classroom. The deadline for submitting the project report is the day of the final exam, which is normally two weeks after the end of the classes.

The PBL is developed in two rounds of four core activities. 
Every student on the team must lead an activity in each round.
The first round is considered completed once all teammates have led one core activity.
All the students will lead another activity in the second round.


Table~\ref{tab:plan} describes, per core activity, the milestone activities for the course and the distribution among homework and classroom sessions.

\begin{table*}[!htp]\centering
\caption{Summary of milestone activities and their description.}\label{tab:plan}
\scriptsize
\begin{tabular}{ccccc}\toprule
Type of session &Session& Core &Activity & Description \\\midrule

\multirow{2}{*}{Problems} & \multirow{2}{*}{P2} & \multirow{4}{*}{CORE 1} & \multirow{2}{*}{Requirements description} & 
\parbox[t]{9cm}{The information system to be modeled is presented. Teams must define the description of requirements.} \\

\multirow{2}{*}{Theory} & \multirow{2}{*}{T3} & & \multirow{2}{*}{Evaluation of requirements} &
\parbox[t]{9cm}{Each team evaluates the description of two other teams. At the end of the activity, a common description is available for all teams.} \\

\midrule

Problems &P3 & \multirow{3}{*}{CORE 2} & Conceptual design &
\parbox[t]{9cm}{Each team develops its conceptual design based on the requirements description.} \\

\multirow{2}{*}{Theory} &\multirow{2}{*}{T4} & & \multirow{2}{*}{Evaluation of conceptual design} &
\parbox[t]{9cm}{Each team evaluates the conceptual design of two other teams. Each team continues with its proposal. Discrepancies are resolved on the blackboard.} \\

\midrule

Problems &P4 &\multirow{2}{*}{CORE 3} & \multirow{2}{*}{Logical design} &
\parbox[t]{9cm}{Each team models the logical design from the conceptual design.} \\

Homework &H1 & & &
\parbox[t]{9cm}{Development of the proposed model in an SW tool.} \\

\midrule

Homework & H2 & \multirow{2}{*}{CORE 4} & Physical design & \parbox[t]{9cm}{The physical part of creating tables starts from the logical design.} \\

Laboratory & L2 & & Database implementation &
\parbox[t]{9cm}{The physical design is implemented in the laboratory. In addition, data must be entered.} \\

\midrule

\multirow{2}{*}{Laboratory} & \multirow{2}{*}{L3} & \multirow{6}{*}{CORE 5} & \multirow{2}{*}{Triggers. Integrity rules} & \parbox[t]{9cm}{In the session, a trigger is created to control project's integrity rules. Each team member prepares a different trigger.} \\

\multirow{2}{*}{Laboratory} & \multirow{2}{*}{L4} & & \multirow{2}{*}{Triggers for audits} &
\parbox[t]{9cm}{An audit trigger is created to keep track of the update operations. Changes to the database introduced by different users must be registered.} \\

\multirow{2}{*}{Laboratory} & \multirow{2}{*}{L5} & & \multirow{2}{*}{Triggers. Business rules} &
\parbox[t]{9cm}{This practice session is focused on the development and implementation through triggers of different business rules of the project. Each team member prepares a different trigger.} \\

\midrule

\multirow{3}{*}{Laboratory} & \multirow{3}{*}{L6} & \multirow{4}{*}{CORE 6}& \multirow{3}{*}{External schemes and views} &
\parbox[t]{9cm}{This practice session is an extension of the activity developed in the sixth problem session. The external schemes and views of the project are implemented. Each team member prepares a different schema.} \\

Laboratory & L7 & & Views and triggers &
\parbox[t]{9cm}{In this practice session, a database update is exercised through views and triggers.} \\

\midrule

\multirow{2}{*}{Laboratory} & \multirow{2}{*}{L8} & \multirow{2}{*}{CORE 7} & \multirow{2}{*}{Triggers and data maintenance} &
\parbox[t]{9cm}{In this practice session, the physical design is modified and an attribute is added to a table. In addition, triggers must be generated to keep it always updated.} \\

\midrule

\multirow{2}{*}{Homework} & \multirow{2}{*}{H3} & \multirow{2}{*}{CORE 8} & \multirow{2}{*}{Final project document} &
\parbox[t]{9cm}{In this activity the team must produce a final document describing all project works performed done during the course. The report includes a backup of the database.} \\

\bottomrule
\end{tabular}
\end{table*}

\subsection{Competencies}

Although evaluating the impact of the TPBL learning methodology on the development of competencies is challenging, the authors believe that some of them have been enhanced. Let us analyze the reasons for these positive effects on the basic and specific competencies under consideration. There are the following basic competencies:

\begin{itemize}
    \item Analysis and synthesis capacity. This competence is ensured as the work team must study the elements of the proposed project and extract the requirements for the database system. Mainly concerned with core activity~1. 
    \item Management of information capacity. This competence is applied when modeling the information system and database of the enterprise proposed in the project.  It is mainly applied in core activities~1 and~2.
    \item Organization and planning skills. The project comprises various activities with several deliverables. The members of each group must plan the different activities and distribute the work for each one. This competence is difficult to apply in subjects based on independent exercises or problems to be performed individually. It is mainly applied in the core activity~1 because students, after defining the project specification, must think about the tasks and timings for its development. In any case, this competence is exercised accross all activities, as the work distribution is required in each of them. 
    \item Autonomous learning capacity. The subject's basic content is taught in the classroom. However, when applying this knowledge to a real situation, doubts and \review{insecurities} emerge. In a long-term project, students must focus more on completing the specific learning gaps adapted to the particular case of the project. It must be considered that students have to evaluate and compare their solutions with those of other groups (core activities~1 and~2) and individual propositions within the team must be debated and integrated into the project (core activities~5-7). 
    \item Problem resolution ability. This basic competence can be applied both in classical teaching and in TPBL methods. The difference is that in the \review{classical teaching} case, students must work more independently and are forced to find practical solutions to the encountered problems. However, in the applied methodology, the solutions of a team will be further compared with those proposed by other groups. This competence is mainly applied in core activities~5-8.
    \item Teamwork capacity, respecting diversity and gender equality. The project is divided into phases or activities. Students must plan and collaborate in all activities, implying that the capacity for teamwork is exercised. Moreover, as previously explained, all members of a group must lead at least two activities (one in each of the two rounds of the project). The activity leader is responsible for assigning tasks to the members of the group, incorporating the solutions proposed by teammates, and ensuring that the activity's objectives are met. 
    The activities are organized in such a way that there is no discrimination in the teams and that all members play all roles.
    \item Writing communication capacity. The use of this competence is ensured by the fact that the last activity, core activity~8, entails drafting a technical report on the project. This document has a bigger scope than the elaboration of independent explanations of different and unconnected exercises. The technical report contains a copy of the implementation of the database, as well as a general description of the project with explanations of the design decisions that have been made.
\end{itemize}

From the specific competencies of the Computer Science degree, the subject must achieve the following:
\begin{itemize}
    \item Capacity to solve problems of information integration. The exercise of this specific competence is ensured as the students must analyze the project description and provide a formal specification on which the project will be based.
    \item Ability to integrate IT solutions and business processes. This is the main goal of the proposed project. This ability is significantly exerted during all project development as the students must arrive at a technological solution for the real practical case presented at the beginning of the project.
    \item Ability to specify, design, implement and maintain IS. These abilities are directly related to the activities to be developed during the project and consequently, students will improve their skills in all these aspects.
  \end{itemize}

\subsection{Students' Evaluation}
During the project development, members of the teams are requested to perform a self-evaluation and to peer-evaluate the performance of their teammates by filling out a questionnaire. This questionnaire is used twice, once at the end of the two phases in which the project's activities are organized. In each round, all members of a team have already assumed the lead role of one activity. 
In addition, in each round, the results of these peer qualifications are sent to the students to get feedback. However, they only receive the average grade assigned by their teammates; they are unaware of the marks awarded by each member of the group. 

Evaluation questionnaires among teammates have been thoroughly described in~\cite{Tomas2021}. These are based on the rubric of~\cite{cinaic2021}, which in turn followed the proposal of~\cite{Chica2011}. 
In these rubrics the following teamwork soft skills are analyzed: 
\begin{itemize}
    \item Contribution (CO): proposition of ideas and improvements.
    \item Participation (PA): involvement in the team's goal.
    \item Manners (M): promotion of team's healthy discussions.
    \item Responsibility (R): compliance with the deadlines.
    \item Attendance (A): do not skip team meetings.
    \item Punctuality (PU): arrival on time to the team meetings.
    \item Conflict resolution (CR): suggestions of new alternatives when disagreements occur in the team.
    \item Leadership (L): fair management of the workload distribution and motivation capability.
\end{itemize}
The questionnaire is presented through Google Forms with the Corubrics tool extension, following the example and recommendations of~\cite{Marques2019}. 

Corubrics supports the rubrics-based evaluation process. Once the form with the rubric is created, it is sent to the team members, so that they can anonymously evaluate their teammates. After completing the forms, Corubrics collects the data in a spreadsheet and optionally presents charts with the results.

As previously mentioned, after each evaluation round, students receive a radar chart showing the marks awarded by the other teammates. 
The second radar chart, corresponding to the second evaluation round, includes the marks of the first evaluation, to highlight their progress. 
Note that there are two implicit evaluations:
\begin{itemize}
    \item the student's perspective compared with the appreciation obtained from the rest of the team; and
    \item the student's evolution compared with their individual evaluated performance.
\end{itemize}

In this regard, students are not only evaluated by the lecturers; instead, the results of the rubrics are used to compute part of their final mark.

\begin{figure*}
     \begin{subfigure}[b]{0.3\textwidth}
        \centering

        \begin{overpic}[trim={3.8cm, 2.28cm, 3.15cm, 2cm}, clip, width=0.5\linewidth, tics=10]
        {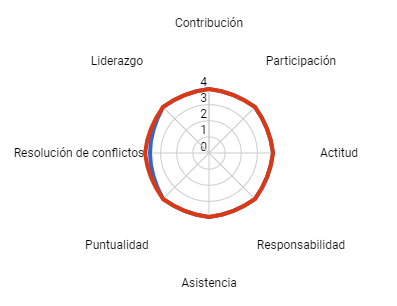}
        \put (40,105) {\Large CO}
        \put (88,85)  {\Large PA}
        \put (100,42) {\Large M}
        \put (88,-2)  {\Large R}
        \put (42,-15) {\Large A}
        \put (-10,-2)  {\Large PU}
        \put (-30,42) {\Large CR}
        \put (-5,85)  {\Large L}
        \end{overpic}
        \vspace{0.5cm}
        \caption{Fair.}
        \label{fig:cat-fair}
     \end{subfigure}
     \hfill
     \begin{subfigure}[b]{0.3\textwidth}
     \centering

        \begin{overpic}[trim={3.8cm, 2.28cm, 3.15cm, 2cm}, clip, width=0.55\linewidth, tics=10]
        {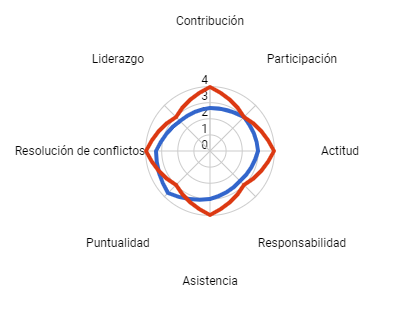}
        \put (40,105) {\Large CO}
        \put (88,85) {\Large PA}
        \put (100,42) {\Large M}
        \put (88,-2) {\Large R}
        \put (42,-15) {\Large A}
        \put (-10,-2) {\Large PU}
        \put (-30,42) {\Large CR}
        \put (-5,85) {\Large L}
        \end{overpic}
        \vspace{0.5cm}
        \caption{Overestimated.}
        \label{fig:cat-oppor}
     \end{subfigure}
     \hfill
     \begin{subfigure}[b]{0.3\textwidth}
     \centering
     \vspace{0.5cm}
        \begin{overpic}[trim={3.84cm, 2.28cm, 3cm, 1.9cm}, clip, width=0.5\linewidth, tics=10]
        {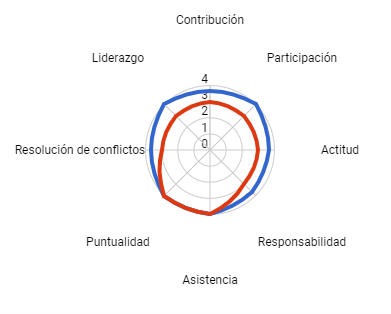}
        \put (40,105) {\Large CO}
        \put (80,85) {\Large PA}
        \put (90,42) {\Large M}
        \put (80,0) {\Large R}
        \put (42,-10) {\Large A}
        \put (-10,-2) {\Large PU}
        \put (-30,42) {\Large CR}
        \put (-5,85) {\Large L}
        \end{overpic}
        \vspace{0.4cm}
        \caption{Underestimated.}
        \label{fig:cat-under}
     \end{subfigure}
        \caption{Examples of radar charts illustrating the different detected behavior. Red lines correspond to the self-evaluation, while blue lines represent the results of the peer evaluation obtained by the student.}\label{fig:radars}
\end{figure*}
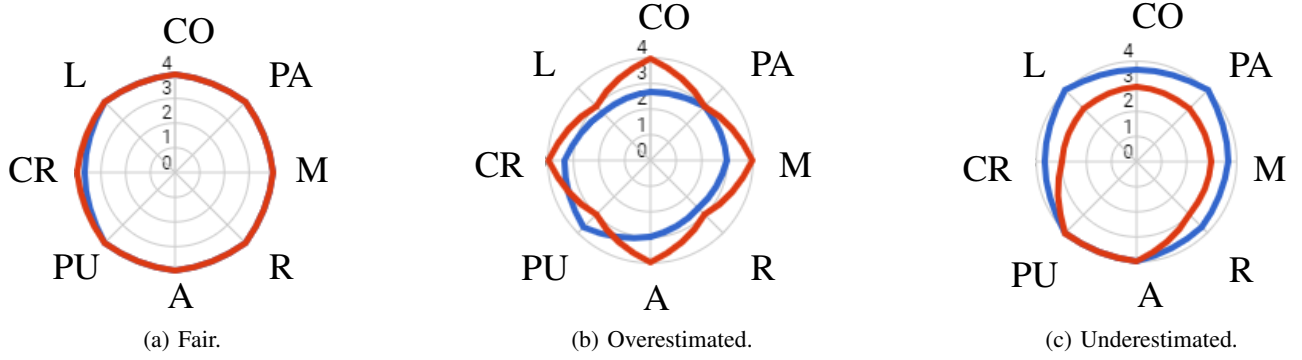

The metrics obtained from the rubrics have been used to measure students' performance within their teams.
Each member is expected to express their appreciations to their teammates.
Rubrics' results are presented as radar charts, allowing students to be categorized.
Figures~\ref{fig:cat-fair}-\ref{fig:cat-under} show examples of these radar charts. The marks range from 0 to 4, with 4 being the maximum grade. Further, the red line represents the self-evaluation, while the blue line represents the average the the rest of the teammates' peer evaluation.

These charts have allowed us to identify three categories of students:
\begin{itemize}
    \item Fair. The team members' appreciation is similar to their self-appreciation. Figure~\ref{fig:cat-fair} shows an example of this category.
    \item Overestimated. Members of the team consider that an individual's performance is lower than their self-appreciation. This behavior may also be understood as optimistic. Figure~\ref{fig:cat-oppor} shows an example of this category.
    \item Underestimated. Members of the team consider that the individual's performance is higher than their self-appreciation. This behavior can indicate that the student undervalues their effort. Figure~\ref{fig:cat-under} shows an example of this category.
\end{itemize}

\section{Results}\label{sec:results}

In the academic year 2021/22, the results of this educational experience have been analyzed from different perspectives: a) the evaluation of the project by the teaching staff and the students' working groups, b) the general opinion of the students about the project and c) the final students' marks.
Further, these results are compared with results obtained from previous years. 

\subsection{Project and Teamwork Evaluation}
As previously explained, each student makes a self-evaluation and is peer-evaluated by their teammates after each of the two rounds of the core activities.
These evaluations provide important insight into students' evolution as the project progresses.

\begin{figure}[htp]
\centering
\includegraphics[trim={3cm, 0.25cm, 2.75cm, 0.5cm}, clip, width=0.8\linewidth]{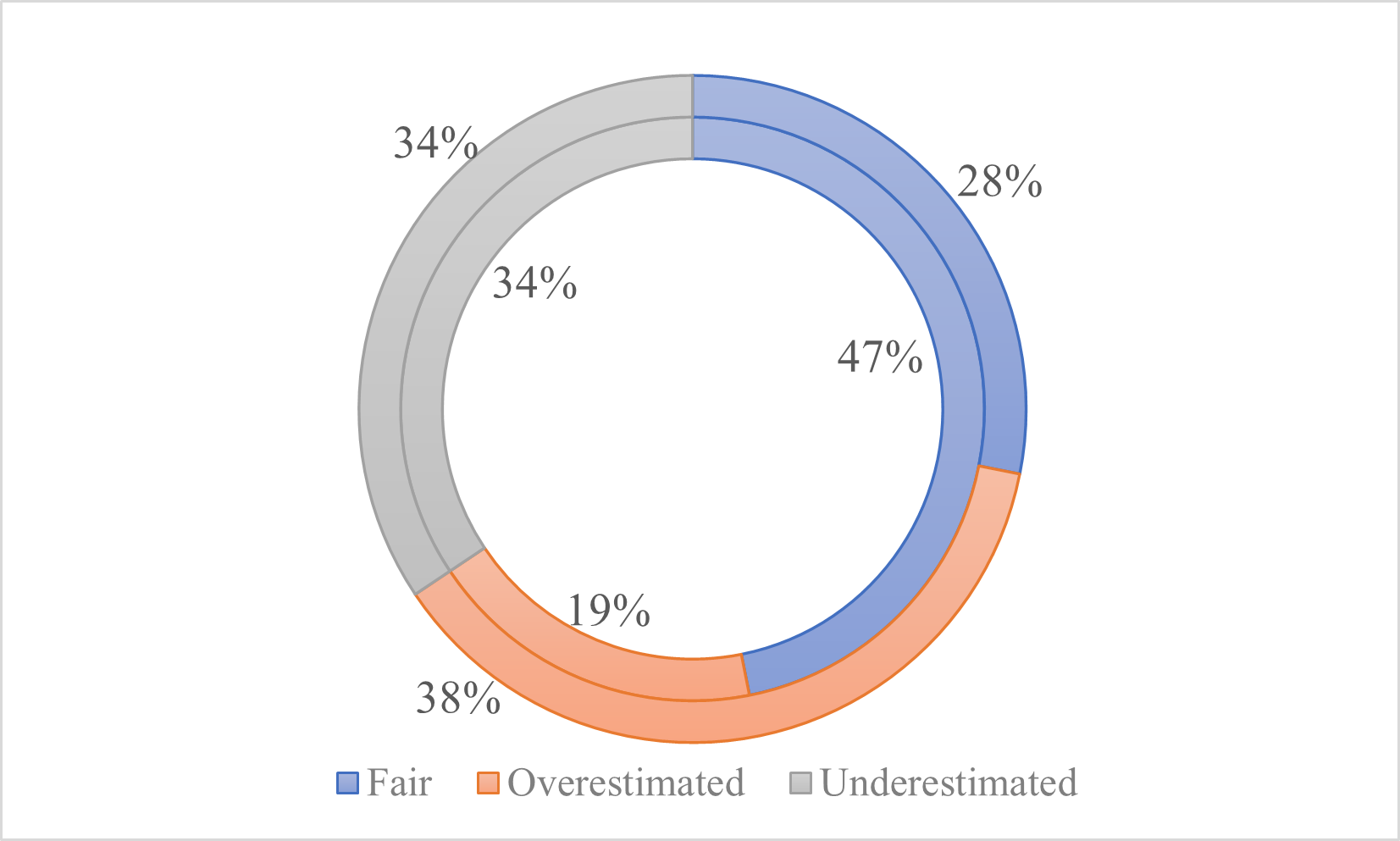}
\caption{Classification of students after comparing their self-evaluation with the peer-evaluation of their teammates. The inner circle corresponds to the \nth{1} round, and the outer circle to the \nth{2}.}
\label{fig:tarta}
\end{figure}

Figure~\ref{fig:tarta} shows the percentage of students classified into the different categories (fair, overestimated, and underestimated) in the two evaluation rounds.
The figure shows an increase in overestimated students at the expense of the fair class.
The differences between the two evaluations could be due to several (and probably interrelated) factors, including the following:

\begin{itemize}
    \item In the first questionnaire, students do not adequately evaluate their mates, assigning them extremely high values, which is difficult to overcome in the second questionnaire.
    \item In the second evaluation round, students have a better understanding of both, their teammates and the questionnaire.
    \item The workload of the students when the second questionnaire is answered is higher, not only due to this subject but also due to the addition of homework from the other subjects. Consequently, some students have reduced their participation in the working group.
\end{itemize}

Table~\ref{tab:detail-auto} contains the aggregated evolution of how the students have evaluated themselves. 
This table classifies the evolution of the self-evaluations into three groups corresponding to students  who have not changed their evaluation (``same''), others who \review{report} a decrease in performance (``worsen''), and finally those who consider that their contribution has increased (``improve'').
Most students evaluated themselves equally in both rounds.
Nevertheless, there are differences in some metrics.
The self-evaluation for the ``contribution metric'' (CO) shows an increase in students who are performing worse.
\review{However, in general, students regard themselves as more punctual (PU) and better leaders (L) once they have finished the project.}

\begin{table}[!ht]\centering
\begin{tabular}{ccccccccc}\toprule
        ~ & CO & PA & M & R & A & PU & CR & L \\ \midrule
        Same & 33 & 32 & 33 & 30 & 36 & 34 & 29 & 32  \\ \midrule
        Worsen & 5 & 4 & 3 & 4 & 2 & 1 & 5 & 2  \\ \midrule
        Improve & 2 & 4 & 4 & 6 & 2 & 5 & 6 & 6  \\ 
        \bottomrule
     \end{tabular}
        \caption{Evolution of self-evaluations.}
    \label{tab:detail-auto}
\end{table}

In the case of the peer-evaluations, Table~\ref{tab:detail-coev} shows how students have evolved from one round of evaluations to another from their teammates' perspectives.
Again, the predominant pattern is to have the same evaluation for the two rounds.
However, the ``same'' rate, in this case, is lower than that in the self-evaluation. 
Teams appreciate greater behavioral diversity among their members. Generally, the team members' performance worsens.

\begin{table}[!ht]\centering
\begin{tabular}{ccccccccc}\toprule
        ~ & CO & PA & M & R & A & PU & CR & L \\ \midrule
        Same & 23 & 23 & 18 & 23 & 26 & 20 & 20 & 24  \\ \midrule
        Worsen & 8 & 11 & 10 & 11 & 11 & 16 & 8 & 10  \\ \midrule
        Improve & 9 & 6 & 12 & 6 & 3 & 4 & 12 & 6  \\
        \bottomrule
     \end{tabular}
        \caption{Evolution of peer-evaluations.}
    \label{tab:detail-coev}
\end{table}

\begin{figure*}[!htb]
\centering
\includegraphics[trim={0.5cm, 0.5cm, 0.4cm, 0.25cm}, clip, width=\linewidth]{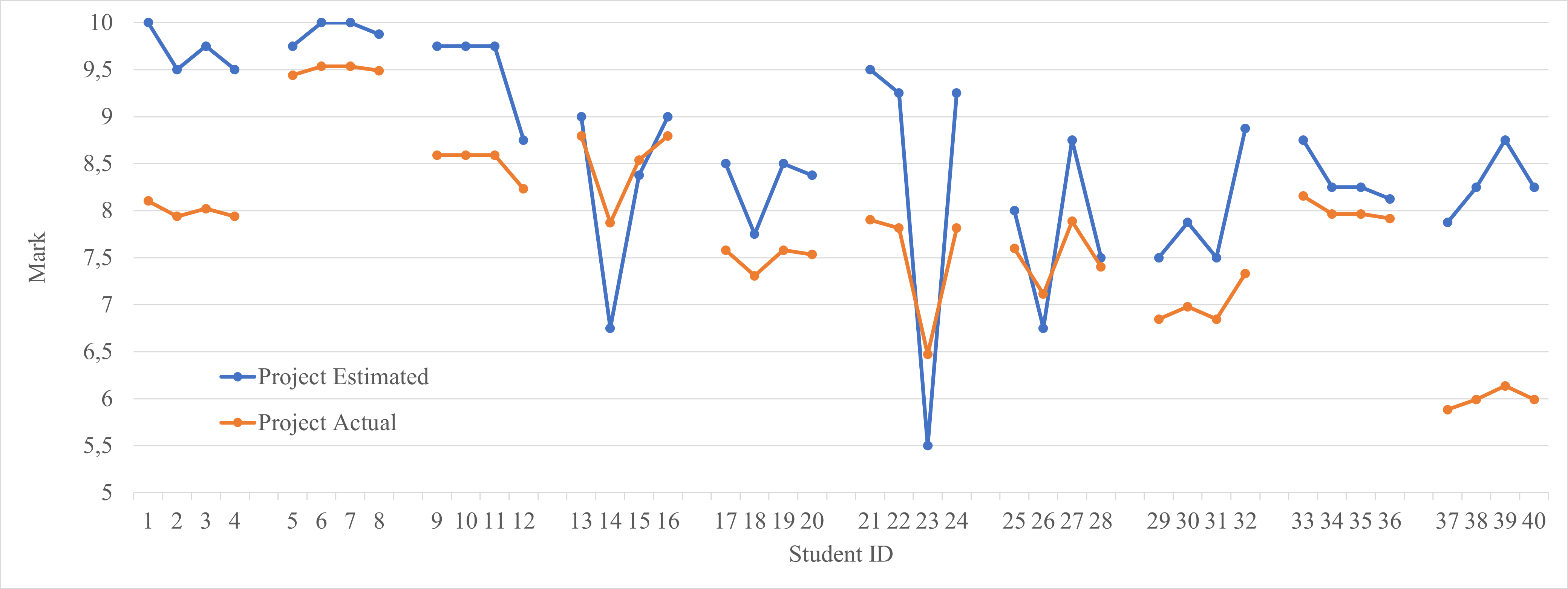}
\caption{Comparison of the estimated and actual marks of the project for the students grouped by their teammates.}
\label{fig:markdiff}
\end{figure*}

The students were asked to estimate their project's final mark in a final form. 
Figure~\ref{fig:markdiff} contains the project mark (in red) and the self-estimated mark (in blue) for each student. In the figure, students are grouped by project teams.

\review{Students expected higher marks than they received. This deviation should be studied in the future since there is not enough information to analyze this fact. However, given that the actual marks follow the patterns of the estimated marks (i.e. they seem to be correlated) a possible cause is a communication problem between the teacher and students about how the evaluation of the project.}

\begin{figure}[!htb]
\centering
\includegraphics[trim={0.5cm, 0.5cm, 0.4cm, 0.25cm}, clip, width=\linewidth]{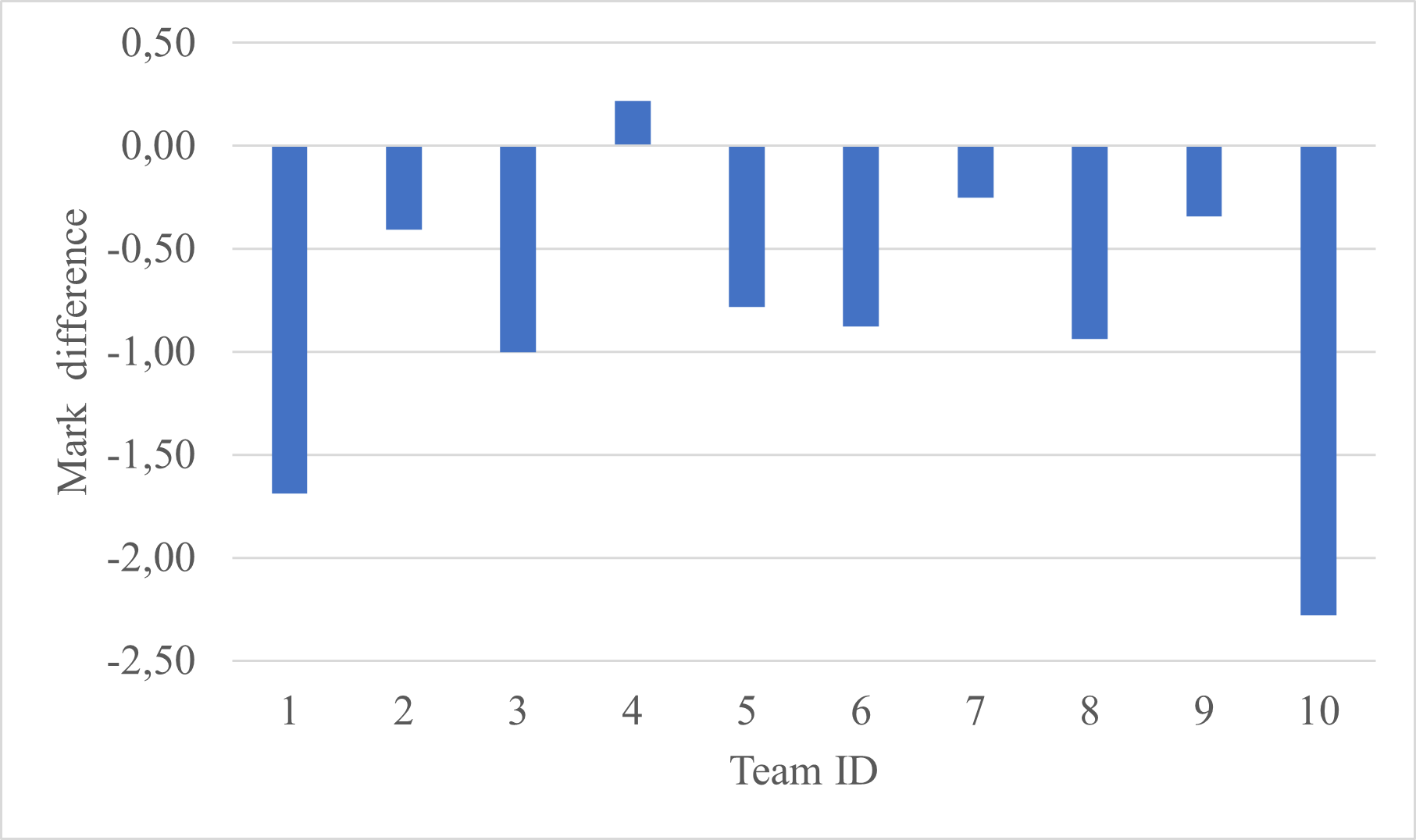}
\caption{Difference between the averaged final and expected project marks per team.}
\label{fig:diffgroup}
\end{figure}

In summary, Figure~\ref{fig:diffgroup} shows the difference between the average actual project mark and the expected average  mark per team. 
\review{Most groups were realistic about their marks since they are within the range of 
$\pm{1}$ point. In particular, group \#1 and \#10 were the less realistic.}

\subsection{Students' Opinion on the Subject and the Project}

During the last four years, students have completed a voluntary questionnaire to determine their opinions on both the subject and the project. This supplementary form is performed at the end of the course after the students have received their grades. 
The questionnaire uses a Likert scale ranging from 0 to 5, with 0 representing the maximum disagreement.

The following are the evaluation questions for the TPBL methodology and other teaching techniques used in the subjects:

\begin{enumerate}
    \item At what point was the subject interesting to you? 
     \item State whether the subject has represented an excessive workload for you outside the classroom (with a score of 5 representing an excessive and unappropriated load, 3 an adequate load, and 0 no load at all).
     \item Rate the methods used in theoretical lectures, which are based on deliverable exercises and peer evaluation.
    \item Rate the method used in practical classes, consisting of a workbook and related exercises.
    \item At what level do you consider that the project has been helpful to acquire better knowledge about the subject's contents?
\end{enumerate}

\begin{figure}[!htb]
\centering
\includegraphics[trim={0.1cm, 0.1cm, 0.5cm, 0.25cm}, clip,width=\linewidth]{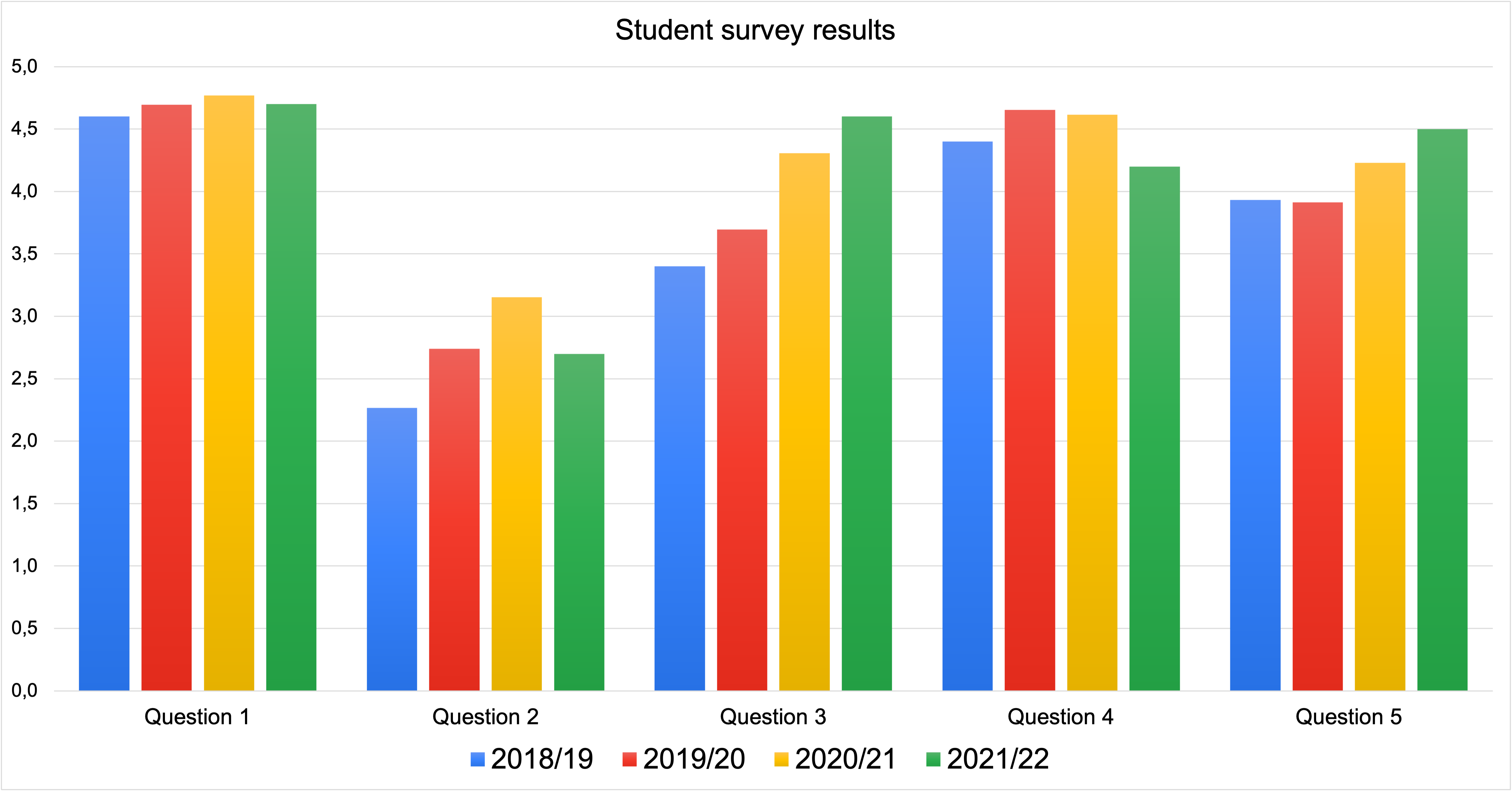}
\caption{Results of the student opinion survey for four academic years.}
\label{fig:studentsopinionsurvey}
\end{figure}

Figure \ref{fig:studentsopinionsurvey} shows the mean values of the above questions. 
The percentage of students who complete this questionnaire varies between 64.5\% and 71\%. 
This figure shows that the students rated the project positively. 
The opinion of the students improves in successive years, probably due to the teachers' accumulated experience in TPBL.

Question 2 is designed to provide an insight into students' perceptions of the workload introduced with the TPBL methodology. 
Figure \ref{fig:studentsopinionsurvey} shows that the perception of the work to be done improves in the first three years. 
Thus, activities outside the classroom were reviewed and simplified. 
In this regard, this value has decreased in the last year.
Nevertheless, students' perception of the workload remains at values close to 3, implying that it does not cost them more than they expectated.

\subsection{Final Results}
Table \ref{tab:finalmarks} presents the final results of the last five academic years, with the percentage of the students passing the subject and the number of enrolled students in the last two columns. TPBL was not introduced as a teaching methodology in the first academic year, 2017/18, and instead, traditional academic work was done. 
In the last academic year, 2021/22, only the marks of the first exam session are included, since the second exam session has not yet been done at the time of writing this paper.

\begin{table}[!ht]\centering
\scriptsize
\begin{tabular}{ccccccc}\toprule
Year & Exam & Laboratory & Project & Final & \%Success & Students \\\midrule
        2017/18 & 5,2 & 8,2 & 7,9 & 7,7 & 65,7 & 35  \\ \midrule
        2018/19 & 6,2 & 8,3 & 8,0 & 8,0 & 78,3 & 25  \\ \midrule
        2019/20 & 6,5 & 8,8 & 8,3 & 8,3 & 89,7 & 29  \\ \midrule
        2020/21 & 6,2 & 8,5 & 8,3 & 8,1 & 79,0 & 38  \\ \midrule
        2021/22 & 6,5 & 7,6 & 7,7 & 8,0 & 80,0 & 40 \\ 
        \bottomrule
     \end{tabular}
\caption{Mean of the final marks by session type. The final mark depends on the weights of each part. The last column corresponds to the enrolled students per year.}
\label{tab:finalmarks}
\end{table}



\review{Table \ref{tab:finalmarks} shows that there has been a slight improvement in the percentage of success since the TPBL methodology was implemented in the academic year 2018/19. This upturn is mainly attributed to better results in the final exam, given that the TPBL methodology help students for a better understanding and application of concepts.}

\section{Conclusions}\label{sec:conclusions}
TPBL makes the learning process more attractive. 
The synergy created by combining teamwork and PBL place students in a real work situation, which a group of people must address a real problem from an organization or institution.

This study presents a PBL methodology applied to advanced database subjects. 
In this experience, a complete information system must be implemented for an entrepreneurial scenario. 
The subjects were comprehensively taught using the TPBL learning technique in the academic year 2021/22, based on the experience obtained in previous years of PBL implementation; all sessions (theory, problems and laboratory) are focused on the problem to be solved by teams of four students.

Thanks to the different types of activities (simple and core), timetable, and teamwork, students learn subject-specific competencies and develop soft skills, such as analysis and synthesis capacity, problem resolution, autonomous learning, critical thinking, and leadership.

The results of the project development have been assessed from three different perspectives: a) peer evaluation within groups and individual self-evaluation were performed using specific rubrics;  b) students’ opinions on the project's impact in their learning outcomes regarding the subject, and c) the subject final results.    

A positive impact was observed with the adoption of the \review{TPBL} methodology.
After using TPBL, final marks were slightly improved and maintained stable at around 8 out of 10. 
Further, the survey to know the student's perception of the use of the TPBL confirms that it helps them acquire and develop new skills. 
Thus, it can be concluded that the project motivates students.


The positive results observed in this teaching experience are a motivation for the ABP2INF\footnote{\url{https://www.uji.es/serveis/use/base/UFIE/innoedu/registreGie/detall?p_grupo_id=157}} innovation teaching group to improve the Computer Science degree by applying the TPBL methodology to other subjects.


However, some issues should be reconsidered in the future. The TPBL methodology is more demanding than classical techniques for teachers' workload. 
When there are more than five teams, it becomes difficult for a single teacher to adequately manage the projects, as meetings and doubt-solving sessions with the teams consume a significant amount of time.
A discoordination and disparity in the development of some activities (particularly core activities 5-7) have also been observed. 
These facts have affected the quality of the final product presented by different teams. 
To address this problem, specific guidelines \review{need to be developed} for the following course.
Finally, delivering the project at the end of the semester, close to the exam period, has a negative effect. 
A way to minimize this drawback is to split the project delivery into two parts: one for the database design (after core activity 4) and the other at the end of the project including the triggers and views for the database, and the final report.

\section*{Acknowledgments}
This work has been developed under the project ``Development of a learning project for subjects EI1038-41 using Project-Based Learning (PBL)'' funded by University Jaume I in the frame of the ``call for grants for educational innovation projects'' for the year 2021 with code 4013/21.
Researcher S.~Iserte is supported by the postdoctoral fellowship APOSTD/2020/026 from Valencian Region Government and European Social Funds.

\bibliographystyle{IEEEtran}
\bibliography{bib}

@inproceedings{Tomas2021,
author = {Tom{\'{a}}s, Vicente R. and Iserte, Sergio and Perez, Miguel},
booktitle = {15th International Technology, Education and Development Conference. INTED},
doi = {978-84-09-27666-0},
pages = {4035--4039},
publisher = {IATED Academy},
title = {{Learning Databases Using Project-based Learning}},
year = {2021}
}

@article{Sulaiman2017,
author = {Sulaiman, Mashitah and {Hj Mat}, Zawiah and {Mod Nizah}, Mohd Aznir and {Abdul Latif}, Latifah},
journal = {Academia.edu},
number = {July 2017},
title = {{The Impact of Teamwork Skills on Students in Malaysian Public Universities}},
url = {The_Impact_of_Teamwork_Skills_on_Students_in_Malaysian_Public_Universities},
year = {2017}
}

@article{Pitsoe2014,
author = {Pitsoe, Victor J. and Isingoma, Peter},
doi = {10.5901/mjss.2014.v5n3p138},
issn = {20392117},
journal = {Mediterranean Journal of Social Sciences},
number = {3},
pages = {138--145},
title = {{How do school management teams experience teamwork: A case study in the schools in the Kamwenge District, Uganda}},
volume = {5},
year = {2014}
}

@article{Berry2001,
author = {Berry, Elizabeth and Lingard, Robert},
doi = {10.18260/1-2--9855},
isbn = {9781612844695},
issn = {01901052},
journal = {ASEE Annual Conference Proceedings},
pages = {9379--9389},
publisher = {IEEE},
title = {{Teaching communication and teamwork in engineering and computer science}},
year = {2001}
}

@techreport{Marques2019,
author = {Marqu{\'{e}}s, Mercedes},
institution = {Universitat Jaume I},
title = {{Dise{\~{n}}o de R{\'{u}}bricas y Evaluaci{\'{o}}n con CoRubrics}},
url = {https://www.uji.es/institucional/estrategia/plans/uji-digital/formacio-digital/cursos/curs-corubrics},
year = {2019}
}

@article{connolly2006constructivist,
author = {Connolly, Thomas M and Begg, Carolyn E},
journal = {Journal of Information Systems Education},
number = {1},
publisher = {Citeseer},
title = {{A Constructivist-Based Approach to Teaching Database Analysis and Design.}},
volume = {17},
year = {2006}
}

@misc{10481/53934,
author = {Rosa-Guillam{\'{o}}n, A and Carrillo-L{\'{o}}pez, P J and Garc{\'{i}}a-Cant{\'{o}}, E},
publisher = {Fern{\'{a}}ndez Revelles, Andr{\'{e}}s B.},
title = {{Learning Based on Projects. A Didactic Experience from the Physical Education Area}},
url = {http://hdl.handle.net/10481/53934},
year = {2019}
}

@article{Oakley04turningstudent,
author = {Oakley, Barbara and Brent, Rebecca and Felder, Richard M and Elhajj, Imad},
journal = {Journal of Student Centered Learning},
pages = {9--34},
title = {{Turning Student Groups into Effective Teams}},
year = {2004}
}

@article{Nattassha2015DatabaseAA,
author = {Nattassha, Ruth and Azizah, Fazat Nur},
journal = {2015 International Conference on Data and Software Engineering (ICoDSE)},
pages = {65--68},
title = {{Database Analysis and Design Learning Tool Based on Problem/Project-based Learning}},
year = {2015}
}

@article{Chica2011,
author = {Chica, Encarnaci{\'{o}}n and Resumen, Merino},
journal = {Escuela Abierta},
pages = {67--81},
title = {{A Proposal for Evaluation Group by Heading}},
url = {http://www.ceuandalucia.es/escuelaabierta/pdf/articulos_ea14pdf/ea14_chica.pdf},
volume = {14},
year = {2011},
month = {nov}
}

@inproceedings{cinaic2021,
author = {Tom{\'{a}}s, Vicente R and Iserte, Sergio and Perez, Miguel and Boronat, Pablo and Castillo, Maribel and Amable, Luis},
booktitle = {Innovaciones docentes en tiempos de pandemia. Actas del VI Congreso Internacional sobre aprendizaje, innovaci{\'{o}}n y cooperaci{\'{o}}n, CINAIC},
doi = {10.26754/uz.978-84-18321-17-7},
isbn = {978-84-18321-17-7},
pages = {256--261},
publisher = {Servicio de Publicaciones. Universidad de Zaragoza.},
title = {{Improving Basic Competences Through Project-based Learning and Teamworking}},
year = {2021}
}

@techreport{Pearlman2000,
author = {Thomas, John W.},
title = {{A Review of Research on Project-based Learning}},
url = {http://www.bie.org/research/study/review_of_project_based_learning_2000},
institution = {Autodesk Foundation},
year = {2000}
}

@TECHREPORT{CS2020,
  author = {The Joint Task Force on Computing Curricula},
  title = {{Computing Curricula 2020. Paradigms for Future Computing Curricula.}},
  institution = {ACM/IEEE Computer Society Press},
  year = {2020},
}

@ARTICLE{Garcia-Robles,
  author={Garcia-Robles, Rocio and Diaz-del-Rio, Fernando and Vicente-Diaz, Saturnino and Linares-Barranco, Alejandro},
  journal={IEEE Transactions on Education}, 
  title={An eLearning Standard Approach for Supporting PBL in Computer Engineering}, 
  year={2009},
  volume={52},
  number={3},
  pages={328-339},
  doi={10.1109/TE.2008.928220}}

@inproceedings{Caceffo,
author = {Caceffo, Ricardo and Gama, Guilherme and Azevedo, Rodolfo},
title = {Exploring Active Learning Approaches to Computer Science Classes},
year = {2018},
isbn = {9781450351034},
publisher = {Association for Computing Machinery},
address = {New York, NY, USA},
url = {https://doi.org/10.1145/3159450.3159585},
doi = {10.1145/3159450.3159585},
booktitle = {Proceedings of the 49th ACM Technical Symposium on Computer Science Education},
pages = {922–927},
numpages = {6},
series = {SIGCSE '18}
}

@article{KUO20191,
title = {Promoting college student’s learning motivation and creativity through a STEM interdisciplinary PBL human-computer interaction system design and development course},
journal = {Thinking Skills and Creativity},
volume = {31},
pages = {1-10},
year = {2019},
issn = {1871-1871},
doi = {https://doi.org/10.1016/j.tsc.2018.09.001},
url = {https://www.sciencedirect.com/science/article/pii/S1871187118301093},
author = {Hsu-Chan Kuo and Yuan-Chi Tseng and Ya-Ting Carolyn Yang},

}

@article{GUO2020101586,
title = {A review of project-based learning in higher education: Student outcomes and measures},
journal = {International Journal of Educational Research},
volume = {102},
pages = {101586},
year = {2020},
issn = {0883-0355},
doi = {https://doi.org/10.1016/j.ijer.2020.101586},
url = {https://www.sciencedirect.com/science/article/pii/S0883035519325704},
author = {Pengyue Guo and Nadira Saab and Lysanne S. Post and Wilfried Admiraal},

}

@article{De_Prada,
title = {Teamwork skills in higher education: is university training contributing to their mastery?.},
journal = {sicol. Refl. Crít},
volume = {35},
year = {2022},
doi = {https://doi.org/10.1186/s41155-022-00207-1},
author = {De Prada, E. and Mareque, M. and Pino-Juste, M.},
}

\end{document}